\documentclass[twocolumn,10pt]{tsfp}
\usepackage{flushend}
\usepackage{graphicx}
\usepackage{amsmath}
\usepackage{amssymb}
\usepackage[authoryear,round]{natbib}
\usepackage{fancyhdr}

\title{Experimental and numerical investigation on preferential alignment of Kolmogorov-size fibers in turbulent channel flow}

\author{\hspace{0in} Eliza Coliban \textsuperscript{\textdagger}, Domenico Zaza and Alfredo Soldati

    \affiliation{

    \hspace{0in} Institute of Fluid Mechanics and Heat Transfer\\

    \hspace{0in} TU Wien\\

    \hspace{0in} Getreidemarkt 9, 1060 Vienna, Austria\\

    \hspace{0in} \textdagger eliza.coliban@tuwien.ac.at

    }    

}

\begin{document}

\maketitle
\thispagestyle{fancy}
\fontsize{9}{11}\selectfont

%%%%%%%%%%%%%%%%%%%%%%%%%%%%%%%%%%%%%%%%%%%%%%%%%%%%%%%%%%%%%%%%%%%%%%
\section*{\fontfamily{phv}\fontsize{9}{11}\bfseries\selectfont ABSTRACT}

We present a combined experimental and numerical investigation of the preferential alignment of Kolmogorov-size, high-aspect-ratio fibers in turbulent channel flow at friction Reynolds numbers $\mathit{Re}_{\tau}=300$ and $550$. Time-resolved volumetric measurements in the TU Wien Turbulent Water Channel are used to simultaneously track fibers and surrounding tracer particles, enabling the reconstruction of fiber trajectories together with a coarse-grained estimate of the local velocity-gradient tensor (VGT). Complementary direct numerical simulations (DNS) of channel flow laden with prolate ellipsoids provide a reference point-particle description. The analysis focuses on the channel core, where the experimental data recover the canonical alignment of vorticity with the intermediate strain-rate eigenvector, thereby supporting the reliability of the reconstructed VGT. We show that fibers preferentially align with the local vorticity direction, while weaker but still non-random alignments are observed with the strain eigenvectors. By measuring finite-time deformation along fiber trajectories through the left Cauchy--Green tensor, we further show that the strongest alignment occurs with the leading principal direction of  Lagrangian stretching. The comparison with DNS shows overall good agreement, while deviations at higher Reynolds number suggest increasing finite-size filtering effects.

\vspace{22pt} 

%%%%%%%%%%%%%%%%%%%%%%%%%%%%%%%%%%%%%%%%%%%%%%%%%%%%%%%%%%%%%%%%%%%%%%
\section*{\fontfamily{phv}\fontsize{9}{11}\bfseries\selectfont INTRODUCTION}
{\fontsize{9}{11}\selectfont

Understanding the orientational and rotational kinematics of small, slender particles in turbulent flows is relevant to a wide range of engineering and environmental processes, including papermaking, fiber suspensions, marine snow formation, and the transport of sediment and microplastics in natural flows \citep{Voth2017}. Beyond these applications, anisotropic tracers such as rods and fibers also provide a direct probe of turbulence itself. When particle inertia is negligible, their rotational kinematics encode statistical information about the local velocity gradient tensor (VGT) sampled along Lagrangian trajectories \citep{Parsa2012,Parsa2014,Brizzolara2021,Zaza_et_al_2026}.  This connection follows directly from Jeffery's equation, which describes the rotation of ellipsoids suspended in Stokesian flows \citep{Jeffery1922}. In the large-aspect-ratio limit, Jeffery kinematics reduce to the evolution equation for the orientation of a material line element in a fluid flow \citep{Dresselhaus1992}, highlighting how a complete understanding of rod alignment in turbulent flows can provide insight into fundamental turbulence mechanisms~\citep{Bentkamp2022}.

In homogeneous isotropic turbulence (HIT), a coherent physical picture has emerged from numerical and experimental studies. Inertialess rods do not remain randomly oriented with respect to the flow, but instead preferentially align with the direction of maximum finite-time Lagrangian stretching~\citep{Ni2014}, while also exhibiting non-trivial alignment with the instantaneous vorticity and with the eigenframe of the instantaneous strain-rate tensor \citep{Pumir2011,Ni2015,Byron2015,Pujara2021}. This preferential orientation has important consequences for fiber rotational statistics, since it modulates tumbling and spinning rates relative to the random-orientation limit \citep{Parsa2014,Byron2015,Pujara2021}; as a result, it also directly impacts measurement techniques based on anisotropic tracers, for which these rotational rates are increasingly used to infer two-point statistics  and velocity-gradient-related quantities~\citep{Brizzolara2021,Zaza_et_al_2026}.

% This preferential orientation has important consequences for rotational statistics, since it modulates tumbling and spinning relative to the random-orientation limit \citep{Parsa2014,Byron2015,Pujara2021}.
%In this sense, rod alignment has become a useful diagnostic for understanding how turbulence organizes anisotropic material elements.

% The picture is more complex in wall-bounded turbulence.  In contrast to HIT, different wall-normal regions are governed by distinct orienting mechanisms: mean shear dominates very near the wall, coherent near-wall structures control alignment throughout the buffer layer, and the outer flow approaches a quasi-isotropic state
% The picture is more complex in wall-bounded turbulence, where particle orientation depend strongly on the distance from the wall. Unlike HIT, different wall-normal regions are associated with distinct orienting mechanisms: mean shear dominates in the immediate vicinity of the wall, coherent near-wall structures govern alignment in the buffer layer, and the outer region progressively recovers a quasi-isotropic state.
% In contrast to HIT, the presence of the wall breaks statistical isotropy and introduces a preferred direction. In addition, the non-homogeneity of the wall-normal direction causes fiber orientation to depend strongly on the wall-normal location. 
The picture is more complex in wall-bounded turbulence, where the mechanisms governing particle orientation depend strongly on the distance from the wall. The relative contributions of mean shear, turbulent fluctuations, and coherent near-wall structures vary across the channel, leading to different preferential orientation  behaviors from the near-wall region to the outer flow~\citep{Cui2021}.
Previous numerical and experimental studies have shown that elongated particles in channels and boundary layers exhibit pronounced streamwise alignment, especially close to the wall \citep{Marchioli2013,Zhao2016,Cui2020,Cui2021,Baker2022,Shaik2023}. Numerical work has further suggested that this behavior should still be interpreted within the Lagrangian stretching-based framework: near-wall fiber orientation is linked to the dominant finite-time Lagrangian stretching direction, which becomes preferentially streamwise in the buffer region \citep{Zhao2016}. 

\begin{figure*}
    \centering
    % Order: Left Bottom Right Top
    \includegraphics[width=6.5in, trim={1cm 3cm 2cm 3cm}, clip]{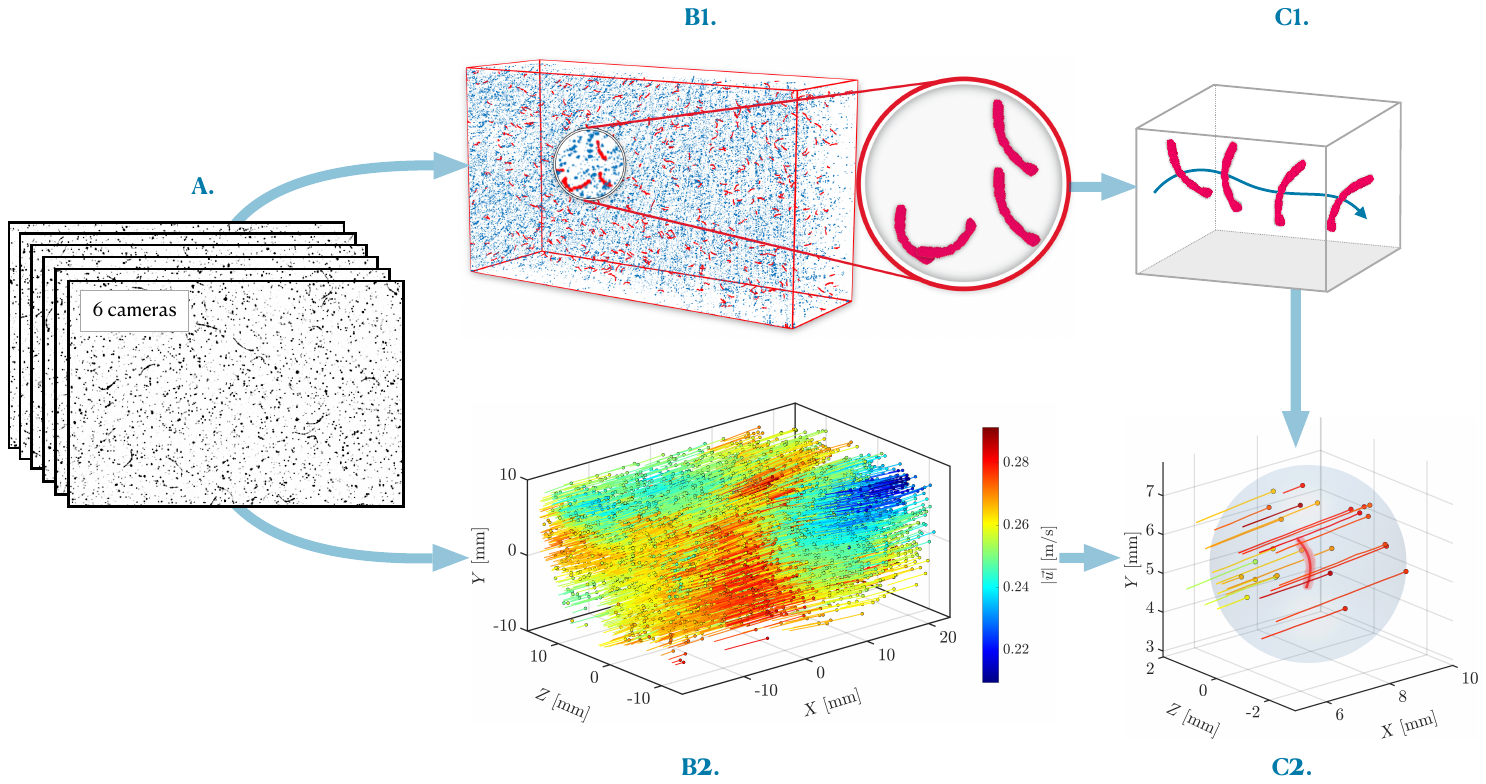}
    \caption{Schematic chart of the methodology of the current study, which includes the following steps: A. Simultaneous acquisition of fibers and flow tracers; B1. Volumetric reconstruction of all objects within the measurement volume, from which the fibers are segmented; C1. Tracking and fitting of fiber objects; B2. Lagrangian (tracer) particle tracking by means of the Shake-The-Box algorithm; C2. Coarse-grain velocity gradient computation centered around the reconstructed fiber objects.}
    \label{Fig1}
\end{figure*}

Yet, simultaneous measurements of rod orientation and the surrounding VGT have so far been achieved experimentally mainly in HIT~\citep{Ni2015}, whereas channel flow experiments have primarily provided statistics of fiber orientation and rotation without directly assessing the combined alignment of vorticity, strain eigenvectors, fiber orientation, and finite-time Lagrangian stretching directions~\citep{Alipour2021,Baker2022,Shaik2023,Giurgiu2024}. In this context, the present work addresses three main questions: (i) do experimental measurements of velocity gradients in the channel core recover the canonical vorticity--strain alignments of turbulence? (ii) do the observed fiber orientations exhibit the expected preferential alignment with the local vorticity and strain eigenframe? and (iii) can the alignment of fibers with the direction of maximum finite-time Lagrangian stretching, so far reported only in DNS, be identified directly in wall-bounded turbulence experiments?

% In this context, the present study addresses three main questions: (i) do experimental measurements of velocity gradients in the channel core recover the canonical vorticity--strain alignments of turbulence? (ii) do the observed fiber orientations exhibit the expected preferential alignment with the local vorticity and strain eigenframe? and (iii) can the alignment of fibers with the direction of maximum finite-time Lagrangian stretching, so far reported only in DNS, be identified directly in wall-bounded turbulence experiments?

To answer these questions, we perform time-resolved volumetric experiments in channel flow, in which slightly curved, nearly neutrally buoyant, high-aspect-ratio fibers are tracked simultaneously with the surrounding tracer field. To the best of our knowledge, this constitutes the first fully three-dimensional simultaneous measurement of both fibers and flow in this configuration, enabling reconstruction of fiber kinematics together with a coarse-grained estimate of the local velocity-gradient tensor along each fiber trajectory. Complementary DNS of passive fibers in the point-particle limit provides a reference description of alignment with local flow gradients. Here we focus on the central region of the channel, where the flow shares important similarities with homogeneous isotropic turbulence (HIT), and use this region as a baseline to assess consistency between experiments and DNS before extending the analysis to the near-wall region in future work.

\vspace{22pt}

%%%%%%%%%%%%%%%%%%%%%%%%%%%%%%%%%%%%%%%%%%%%%%%%%%%%%%%%%%%%%%%%%%%%%%
\subsection*{\fontfamily{phv}\fontsize{9}{11}\bfseries\selectfont Methodology}

The experimental campaign couples time-resolved volumetric measurements of the fluid with Lagrangian tracking of fibers. Measurements were performed in the TU Wien Turbulent Water Channel \citep{Giurgiu2023} at two friction Reynolds numbers, $\mathit{Re}_{\tau}=u_{\tau}h/\nu=300$ and $550$, where $u_{\tau}$ is the friction velocity, $h$ is the channel half-height ($40 \text{ mm}$), and $\nu$ is the fluid kinematic viscosity. The measurement volume is centered at $y = 37 \text{ mm}$ and spans $42 \times 18 \times 28 \text{ mm}^3$ within the laboratory reference frame ($x,y,z$), where $x, y,$ and $z$ denote the streamwise, wall-normal, and spanwise directions, respectively. 

The flow was seeded with nearly neutrally buoyant fibers ($\rho_{f}=1.15 \text{ g/cm}^3$) with a nominal length $\ell=1.2  \ \text{mm}$ and diameter $d=10 \ \mu \text{m}$, as well as $20\ \mu \text{m}$ polyamide tracer particles (LaVision). As demonstrated by \cite{Giurgiu2024} and \cite{Alipour2022}, under the present flow conditions, the fibers behave effectively as rigid translational and rotational tracers. Additionally, the suspension is maintained in the dilute regime, with a fiber volume fraction on the order of $\mathcal{O}(10^{-7})$. At this concentration, the system is characterized by one-way coupling, ensuring that the fibers behave as passive elements that do not influence the velocity field or the underlying turbulent structures. 

Both fibers and tracers were captured simultaneously at a resolution of $56.59 \text{ px/mm}$ by six high-speed cameras arranged in a circular configuration below the measurement volume \citep{Giurgiu2025}. The volume was illuminated by a light sheet generated with a high-speed, dual-cavity Nd:YAG laser ($532 \text{ nm}$). The main flow and acquisition parameters for the two investigated cases are summarized in Table~\ref{Table1}. 

\begin{table}[ht]
\centering
\fontfamily{ptm}\fontsize{9}{9}\selectfont
\caption{Flow and acquisition parameters for the current investigation: Friction Reynolds number  $\mathit{Re}_\tau$, friction velocity $u_\tau$, acquisition frequency $f$, and water kinematic viscosity $\nu$.}
\label{Table1}
\begin{tabular}{c c c c}
\hline\hline
$Re_{\tau}$ & $u_{\tau}$ (mm/s) & $f$ (1/s) & $\nu$ (m$^2$/s) \\
\hline
300 & 7.0 & 500 & $9.39 \times 10^{-7}$ \\
550 & 13.0 & 840 & $9.42 \times 10^{-7}$ \\
\hline\hline
\end{tabular}
\end{table}

 Figure \ref{Fig1} illustrates the experimental data processing workflow. Steps A through B were conducted using the Davis 10 environment, while subsequent steps (C and beyond) were implemented in MATLAB. Panel A presents the raw images of the fibers and tracers, which underwent preprocessing to eliminate measurement noise and enhance the contrast between the two particle types. These images also formed the basis for the Volume Self-Calibration (VSC); by masking the fibers, the tracer-only images were utilized to refine the initial geometric calibration.

For the fiber processing, the present study employs the methodology described in \cite{Giurgiu2024}. The images are processed using tomographic reconstruction based on the Multiplicative Algebraic Reconstruction Technique (MART), enabling a full voxelized representation of both fibers and tracers (see panel B1). From this volume, the fiber objects are segmented, and a polynomial fit is applied to determine their centerlines, curvatures, and principal axes. This procedure facilitates the accurate tracking of the fiber center of mass and orientation over time (see panel C1), providing the complete set of rotational degrees of freedom alongside their spatial trajectories. The principal axes of inertia for each fiber are defined as: $\hat{e}_1$, the axis along the fiber length (hereafter referred to as $\mathbf{p}$), $\hat{e}_2$, the axis within the plane of curvature, and $\hat{e}_3$, the axis perpendicular to the plane of curvature.

To obtain the time-resolved, 3D velocity field, we utilize a coarse-grained velocity gradient tensor estimation based on tracer trajectories recovered via the Shake-The-Box (STB; \cite{Schanz2016}) algorithm (see panel B2). The tracer concentration was increased incrementally to the maximum achievable density. In such experiments, a fundamental trade-off exists: higher tracer concentrations inevitably reduce fiber visibility. Consequently, the seeding density was optimized to the point where fiber light intensities began to diminish significantly relative to the tracer particles. For the final processing, the STB algorithm successfully reconstructed approximately $8,000-10,000$ tracer trajectories for the $Re_\tau = 300$ case, and $7,000-8,000$ trajectories for the $Re_\tau = 550$ case. Relative to the size of the measurement volume, this corresponds to a volumetric seeding density of $0.33-0.47$ particles/mm$^3$.

% In a next step the information of the fiber trajectories (C1) is combined with the lagrangian information of the tracers (C2).

In a next step, the information from the fiber trajectories (C1) is combined with the Lagrangian information from the tracers (C2). The local velocity gradient tensor at the fiber position is estimated from the velocities of nearby tracer particles following the procedure of \citet{Ni2015}. At each time step, all tracer particles within a sphere of radius $R \approx 6\,\eta$ centered on the fiber centroid are identified. For $Re_\tau = 300$ this corresponds to $R = 3.51~\mathrm{mm}$, and for $Re_\tau = 550$ to $R = 2.32~\mathrm{mm}$. Their velocities are modelled by a first-order Taylor expansion of the velocity field about the centroid of the tracer cloud:
\begin{equation}
    \Delta u_i^{(n)} = \tilde{A}_{ij}\,\Delta x_j^{(n)},
    \label{eq:gradient_model}
\end{equation}
where $\Delta u_i^{(n)} = u_i^{(n)} - \bar{u}_i$  and  $\Delta x_j^{(n)} = x_j^{(n)} - \bar{x}_j$,  with  $\bar{u}_i$ and $\bar{x}_j$ denoting the mean tracer velocity and mean tracer position within the interrogation volume. Here, $n=1,\dots,N$ labels the tracers within the sphere, while $i,j=1,2,3$ denote Cartesian components. The coarse-grained velocity gradient tensor $\tilde{A}_{ij}$ is obtained by minimizing the sum of squared residuals between the measured and modelled tracer velocities. Writing the centered tracer positions and velocities as the matrices $\mathbf{X}\in\mathbb{R}^{N\times 3}$ and $\mathbf{U}\in\mathbb{R}^{N\times 3}$, the least-squares problem reads $\mathbf{U} = \mathbf{X}\,\tilde{\mathbf{A}}^{\top}$, so that $\tilde{\mathbf{A}}^{\top}
=
(\mathbf{X}^{\top}\mathbf{X})^{-1}\mathbf{X}^{\top}\mathbf{U}$.

Estimates of the VGT that fail the quality criteria employed in  \citet{Ni2015}---minimum tracer count within the considered sphere, spatial isotropy of the tracer cloud, proximity of the fiber to the tracer centroid, and relative incompressibility---are discarded. The resulting $\tilde{A}_{ij}$ is a spatially filtered representation of the velocity gradient, with an effective resolution set by the interrogation radius $R$. Its symmetric and skew-symmetric parts, $\tilde{S}_{ij} = \tfrac{1}{2}(\tilde{A}_{ij} + \tilde{A}_{ji})$ and $\tilde{\Omega}_{ij} = \tfrac{1}{2}(\tilde{A}_{ij} - \tilde{A}_{ji})$, represent the coarse-grained strain-rate and rotation-rate tensors at the scale of the interrogation volume.

To complement the experiments, we also conduct direct numerical simulations (DNS) of turbulent channel flow transporting elongated particles at matching friction Reynolds numbers, i.e, $Re_{\tau}=300$ and $550$. In these simulations, the Navier–Stokes equations are solved in a plane channel domain with dimensions $L_x = 4\pi h$, $L_y = 2h$, and $L_z = 2\pi h$ along the streamwise ($x$), wall-normal ($y$), and spanwise ($z$) directions, respectively. 
The numerical approach follows established pseudo-spectral methods, already employed in~\cite{Zaza_2024,Zaza_et_al_2026}. Specifically, a Fourier–Galerkin discretization is employed in the periodic streamwise and spanwise directions, while a Chebyshev–tau formulation is used in the wall-normal direction~\citep{Kim_Moin_Moser_1987}.  Time advancement is carried out using a semi-implicit scheme, where viscous terms are treated with the Crank–Nicolson method and nonlinear convective terms with a two-step Adams–Bashforth scheme. This results in a second-order accurate temporal discretization. The spatial resolution and numerical parameters adopted in the simulations are summarised in Table~\ref{TableDNS}. 

\begin{table}[ht]
\centering
\fontfamily{ptm}\fontsize{9}{9}\selectfont
\caption{Summary of the direct numerical simulation (DNS) parameters: Shear Reynolds number $\mathit{Re}_{\tau}$, number of grid points in the streamwise ($N_x$), wall-normal ($N_y$), and spanwise ($N_z$) directions, and corresponding grid spacings in wall units, $\Delta x^+$ and $\Delta z^+$. The wall-normal resolution is reported in terms of the minimum and maximum spacing in wall units, $\Delta y^+_{w}$ and $\Delta y^+_{c}$, associated with the near-wall region and the channel centerline, respectively. %The wall-normal discretization employs Chebyshev–Gauss–Lobatto points, leading to a non-uniform grid in $y$.
}

\label{TableDNS}
\begin{tabular}{c c c c c c c c }
\hline\hline
$\mathit{Re}_{\tau}$ & $N_x$ & $N_y$ & $N_z$ & $\Delta x^+$ & $\Delta z^+$ & $\Delta y^+_{w}$ & $\Delta y^+_{c}$  \\
\hline
300 & 768 & 301 & 384 & 7.36 & 7.36 & 0.016 & 3.14   \\
550 & 1024 & 385 & 768 & 10.1 & 6.75 & 0.019 & 4.50  \\
\hline\hline
\end{tabular}
\end{table}

The fibers in the simulation are treated as elongated, neutrally buoyant, and axisymmetric ellipsoids, and their evolution is tracked using the Lagrangian approach within the point-particle framework~\citep{Voth2017}. Particle translational velocity is set equal to the velocity of the fluid interpolated at the location of the particle, while the fiber orientation evolves according to the Jeffery equation~\citep{Jeffery1922}. Following the methodology proposed by~\cite{Pujara2021} and corroborated by the recent work of~\cite{Zaza_et_al_2026}, the strain-rate and rotation-rate tensors entering the Jeffery dynamics are not obtained from the local velocity gradient, but from a velocity gradient tensor filtered at the scale of the experimental fiber length. In practice, the instantaneous fluid velocity field is convolved with an isotropic Gaussian kernel whose characteristic width matches the experimental fiber length, and spatial derivatives are then evaluated from the filtered field. The resulting filtered strain-rate tensor $\overline{\mathbf{S}}^{\ell}$ and vorticity vector $\overline{\boldsymbol{\omega}}^{\ell}$ are interpolated at the particle position, and the instantaneous fiber angular velocity is then directly prescribed as
\begin{equation}
\boldsymbol{\omega}_{p}
=
\frac{1}{2}\overline{\boldsymbol{\omega}}^{\ell}
+
\frac{\alpha^{2}-1}{\alpha^{2}+1} \,\mathbf{p}\times\left(\overline{\mathbf{S}}^{\ell}\mathbf{p}\right),
% \qquad
% \lambda=\frac{\alpha^{2}-1}{\alpha^{2}+1},
\end{equation}
where $\mathbf{p}$ is the unit vector aligned with the ellipsoid symmetry axis and $\alpha$ is the particle aspect ratio, which is set to $\alpha=120$ in agreement with the experimental fibers.

% The fibers in the simulation are treated as elongated, neutrally buoyant, and axisymmetric ellipsoids, and their evolution is tracked using the Lagrangian approach within the point-particle framework. We consider the point-particle limit for the fibers hydrodynamics: each fiber is assumed to be small enough that it samples a single-point velocity gradient at its center location. Under these conditions, the fiber’s orientation obeys Equation 1.

% \emph{DNS---}  For all considered shear Reynolds numbers ($\mathit{Re}_\tau = 180$, $360$, $700$), the grid spacing in wall units is maintained at approximately $\Delta x^+ = \Delta z^+ \approx 5.9$ in the homogeneous directions. In the wall-normal direction, the resolution varies from $\Delta y^+ = 0.02$ up to $2.82$ for $\mathit{Re}_\tau = 180$, up to $3.77$ for $\mathit{Re}_\tau = 360$, and up to $5.24$ for $\mathit{Re}_\tau = 700$.
%  Periodic boundary conditions are applied along $x$ and $z$, and no-slip conditions are imposed at the walls located at $y = \pm h$.

% A pressure-eliminated formulation of the Navier–Stokes equations is adopted.

\subsection*{\fontfamily{phv}\fontsize{9}{11}\bfseries\selectfont Results and Discussions}

As a validation of the present velocity gradient measurements, we examine the relative orientation between the experimental vorticity vector, whose direction is denoted by $\mathbf{e}_{\boldsymbol{\omega}}$, and the eigenvectors of the  strain-rate tensor, which are denoted by $\mathbf{e}_1$, $\mathbf{e}_2$, and $\mathbf{e}_3$ and correspond, respectively, to the most extensive, intermediate, and most compressive principal strain directions.  %, associated with the local Eulerian stretching and compression of fluid elements. 
This provides a robust test of the measured velocity gradients, since HIT consistently exhibits characteristic alignments between the rotational and straining components of the VGT, whereby vorticity preferentially aligns with $\mathbf{e}_2$, shows weak alignments with $\mathbf{e}_1$, and tends to lie perpendicular to $\mathbf{e}_3$~\citep{Ashurst1987,Ni2015}.
\begin{figure}[ht]
    \centering
    \includegraphics[width=1.0\columnwidth]{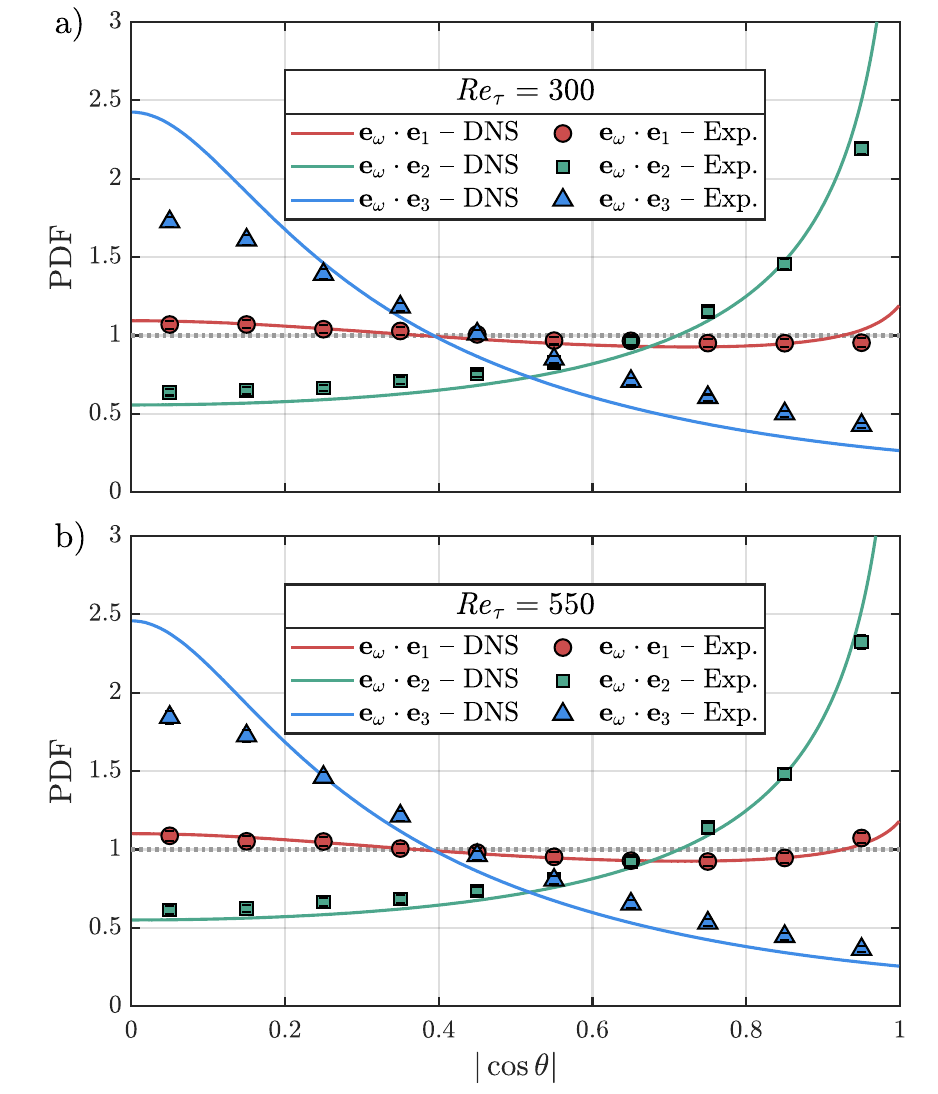}
% \caption{
% Probability density functions of the cosine of the angle  between the vorticity vector, $\omega$, and the eigenvectors $e_1$, $e_2$, and $e_3$ of the strain-rate tensor for (a) $Re_{\tau} = 300$ and (b) $Re_{\tau} = 550$. The dashed line indicates the distribution for random orientation.}
\caption{Probability density functions of $|\cos\theta|$, where $\theta$ denotes, depending on the eigenvector considered, the angle between the vorticity vector $\boldsymbol{\omega}$ and the eigenvectors $\mathbf{e}_1$, $\mathbf{e}_2$, and $\mathbf{e}_3$ of the strain-rate tensor, for (a) $\mathit{Re}_{\tau}=300$ and (b) $\mathit{Re}_{\tau}=550$. The eigenvectors $\mathbf{e}_1$, $\mathbf{e}_2$, and $\mathbf{e}_3$ correspond to the most extensive, intermediate, and most compressive principal strain directions, respectively. Solid lines denote present DNS results, conditioned on the central third of the channel, while  
%corresponding to $200 \leq y^+ \leq 300$ for $\mathit{Re}_{\tau}=300$ and $365 \leq y^+ \leq 550$ for $\mathit{Re}_{\tau}=550$; 
symbols denote experimental results obtained for the central region of the channel. %The dotted line at $\mathrm{PDF}=1$ corresponds to the uniform distribution of $|\cos\theta|$ expected for randomly oriented vectors.
Error-bars represent confidence intervals at 99\% confidence level and have been computed using the bootstrap-based method.
}
\label{figure2}
\end{figure}

 In this vein, Figure~\ref{figure2} reports the probability density functions of $|\cos\theta|$, where $\theta$ denotes, successively, the angle between ${\boldsymbol{\omega}}$ and each principal strain direction. Experimental results are compared with the corresponding DNS data at the two considered Reynolds numbers for the central region of the channel. Specifically, the experimental measurements were acquired over the wall-normal intervals $ 206\lesssim y^+ \lesssim 300$ for $\mathit{Re}_{\tau}=300$ and $ 378\lesssim y^+ \lesssim 550$ for $Re_{\tau}=550$, with $y^+=yu_\tau/\nu$ wall-normal coordinate expressed in wall units . Likewise, DNS statistics are conditioned on the central third of the channel, corresponding to $200 \leq y^+ \leq 300$ for $\mathit{Re}_{\tau}=300$ and $365 \leq y^+ \leq 550$ for $\mathit{Re}_{\tau}=550$.

For both Reynolds numbers considered, the measurements in Figure~\ref{figure2} clearly recover the canonical HIT alignment statistics of the VGT. In particular, the probability density associated with $|\mathbf{e}_{\boldsymbol{\omega}}\!\cdot\!\mathbf{e}_2|$ increases monotonically and exhibits a pronounced peak near unity, indicating the strong preferential alignment between vorticity and the intermediate strain-rate eigenvector $\mathbf{e}_2$. Conversely, the distribution of $|\mathbf{e}_{\boldsymbol{\omega}}\!\cdot\!\mathbf{e}_3|$ is largest near zero and decreases steadily, reflecting the tendency of vorticity to remain perpendicular to the most compressive strain direction. The PDF relative to $|\mathbf{e}_{\boldsymbol{\omega}}\!\cdot\!\mathbf{e}_1|$ remains comparatively flat and close to the uniform distribution indicating no preferential orientation of $\boldsymbol{\omega}$ with respect to $\mathbf{e}_1$.
%The comparison with DNS results shows very good agreement, reproducing both the overall trends and the quantitative distributions. In particular, the monotonic growth of the $\mathbf{e}_2$ curve, the decay of the $\mathbf{e}_3$ curve, and the nearly flat behaviour of the $\mathbf{e}_1$ statistics are all reproduced by the experiments. Modest deviations appear only near the extrema, where statistical convergence is weaker and measurement uncertainties are expected to be more significant. 
%The comparison with DNS provides further confidence in the accuracy of the reconstructed velocity gradients and in the robustness of the measurement methodology adopted in the present work.
The comparison with DNS results shows very good agreement, reproducing both the overall trends and the quantitative distributions. Modest deviations appear only near the extrema for $|\mathbf{e}_{\boldsymbol{\omega}}\!\cdot\!\mathbf{e}_3|$.

Having shown that the reconstructed velocity gradients capture the expected alignment statistics of HIT, we now investigate how fibres orient with respect to the components of the local VGT. Figure~\ref{figure3} reports the probability density functions of $|\cos\theta|$, where $\theta$ now denotes, successively, the angle between the fibre orientation vector $\mathbf{p}$ and the vorticity direction $\mathbf{e}_{\boldsymbol{\omega}}$, and between $\mathbf{p}$ and the strain-rate eigenvectors $\mathbf{e}_1$, $\mathbf{e}_2$, and $\mathbf{e}_3$. Experimental results are compared with the corresponding DNS data in the same central-channel region considered in Figure~\ref{figure2}. 
\begin{figure}[ht]
    \centering
    \includegraphics[width=1.0\columnwidth]{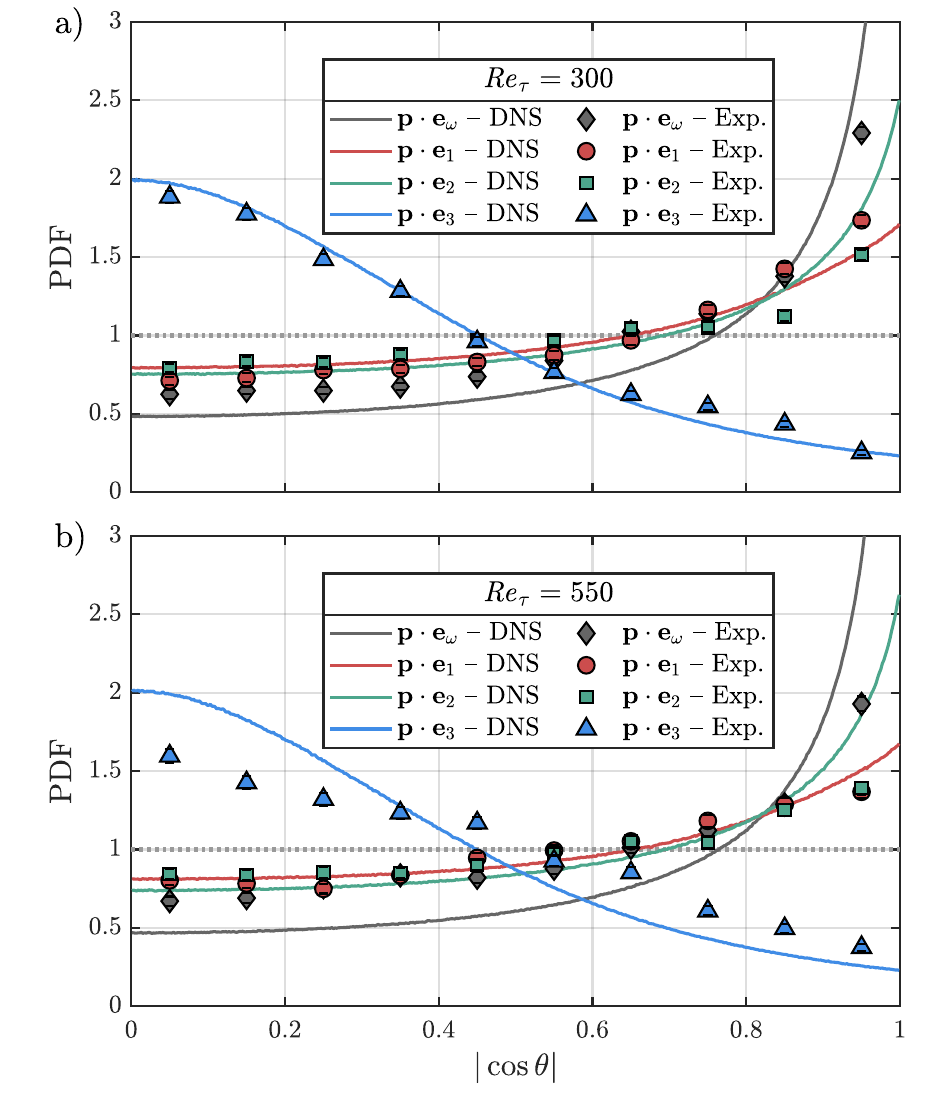}
\caption{Probability density functions of $|\cos\theta|$, where $\theta$ denotes, depending on the vector considered, the angle between the fiber orientation vector $\mathbf{p}$ and the vorticity vector $\boldsymbol{\omega}$, and the eigenvectors $\mathbf{e}_1$, $\mathbf{e}_2$, and $\mathbf{e}_3$ of the strain-rate tensor, for (a) $\mathit{Re}_{\tau}=300$ and (b) $\mathit{Re}_{\tau}=550$. Solid lines denote present DNS results and symbols denote experimental results obtained for the central region of the channel. Error-bars represent confidence intervals at 99\% confidence level and have been computed using the bootstrap-based method.}
    \label{figure3}
\end{figure}

For both Reynolds numbers considered, the numerical and experimental distributions in Figure~\ref{figure3} show that fibers are not randomly oriented with respect to the components of the local VGT. The strongest preferential alignment is consistently observed with the vorticity direction $\mathbf{e}_{\boldsymbol{\omega}}$, whose PDF increases markedly toward $|\cos\theta|=1$ in both experiments and DNS. This trend is in agreement with the measurements and simulations of \citet{Ni2015} in HIT, where rods were likewise found to align more strongly with vorticity. Alignment with the strain eigenvectors is weaker: the distributions associated with $\mathbf{e}_1$ and $\mathbf{e}_2$ still depart from the random reference level, and  alignment with the most compressive direction $\mathbf{e}_3$ is strongly suppressed, with the highest probability occurring near $|\cos\theta|=0$.

% For both Reynolds numbers considered, the numerical and experimental distributions in Figure~\ref{figure3} show that fibers are not randomly oriented with respect to the components of the local VGT. The strongest preferential alignment is consistently observed with the vorticity direction $\mathbf{e}_{\boldsymbol{\omega}}$, whose PDF increases markedly toward $|\cos\theta|=1$ in both experiments and DNS. This trend is in agreement with the measurements and simulations of \citet{Ni2015} in HIT, where rods were likewise found to align more strongly with vorticity.
% Consistently, in the present measurements, the probability distributions associated with $\mathbf{e}_1$ and $\mathbf{e}_2$ exhibit clear but weaker preferential alignment, while alignment with the most compressive direction $\mathbf{e}_3$ is strongly suppressed, with the highest probability occurring near $|\cos\theta|=0$.

A more subtle feature concerns the relative ordering of the alignments with $\mathbf{e}_1$ and $\mathbf{e}_2$ at the two considered $\mathit{Re}_\tau$. At $\mathit{Re}_{\tau}=300$ (Fig.~\ref{figure3}a), the experimental data show slightly stronger alignment with $\mathbf{e}_1$, whereas Jeffery's model in the DNS predicts a stronger alignment with $\mathbf{e}_2$. A similar experiment--simulation discrepancy was already reported by \citet{Ni2015} in HIT, where the exchanged ordering of the two statistics was attributed to their sensitivity to Reynolds number and to inevitable differences in turbulence forcing between the experimental flow configuration and direct numerical simulation. At the higher Reynolds number, $\mathit{Re}_{\tau}=550$ (Fig.~\ref{figure3}b), the preferential alignments are generally weakened relative to DNS, and the statistics associated with $\mathbf{e}_1$ and $\mathbf{e}_2$ become nearly indistinguishable. Overall, all alignment signatures appear attenuated relative to the lower-Reynolds-number case, consistent with finite-size filtering effects not fully captured by the simulated point-fibers sampling the filtered VGT. As the ratio between fiber length and Kolmogorov scale increases, the fibers sample a broader range of velocity gradients along their extent, thereby line-averaging the small-scale fluctuations below their length. Thus, the residual discrepancies observed with DNS results suggest that, despite capturing the leading trends, the simple filtered Jeffery model may not fully represent the scale-dependent alignments.

The alignment trends reported so far are consistent with previous studies indicating that fiber orientation is governed by the cumulative fluid deformation experienced along trajectories rather than by the instantaneous  Eulerian strain field. In particular, both rods and vorticity were shown to preferentially align with the dominant Lagrangian finite-time stretching direction~\citep{Ni2014,Ni2015}. Moreover, combined experimental and numerical results by \citet{Xu2011} demonstrated that the classical instantaneous alignment of vorticity with the intermediate strain eigenvector does not reflect a static geometrical constraint, but rather coexists with a short-time dynamical tendency of vorticity to rotate toward the initially most extensional stretching direction. Building on these observations, Figure~\ref{figure4} examines the alignment of fibers with the finite-time Lagrangian stretching directions, defined by the eigenvectors of the left Cauchy--Green deformation tensor, $\mathbf{C}^{(L)}=\mathbf{F}\mathbf{F}^{T}$, 
where $\mathbf{F}$ is the deformation-gradient tensor accumulated over a finite time interval $\Delta t$. It is obtained by integrating along the tracer-fiber trajectory the evolution equation: $
{\mathrm{d}\mathbf{F}}/{\mathrm{d}t}=\mathbf{A}(t)\mathbf{F}$, with initial condition $\mathbf{F}(t_0)=\mathbf{I}$. Here,  $\mathbf{A}(t)$ is the measured VGT sampled by the fiber and $\mathbf{I}$ is the identity tensor. Due to the finite duration over which continuous gradient measurements can be accurately obtained along experimental trajectories, the integration is performed over a relatively short time window, here chosen as $\Delta t=0.10\,\tau_{\eta}$, where $\tau_{\eta}$ is the Kolmogorov time scale, defined as $\tau_{\eta}=(\nu/\epsilon)^{1/2}$, with $\epsilon$ denoting the mean turbulent dissipation rate. In the present experiments, we obtain $\tau_{\eta}=0.408\,\mathrm{s}$ for $\mathit{Re}_{\tau}=300$ and $\tau_{\eta}=0.149\,\mathrm{s}$ for $\mathit{Re}_{\tau}=550$.

\begin{figure}[ht]
    \centering
    \includegraphics[width=1.0\columnwidth]{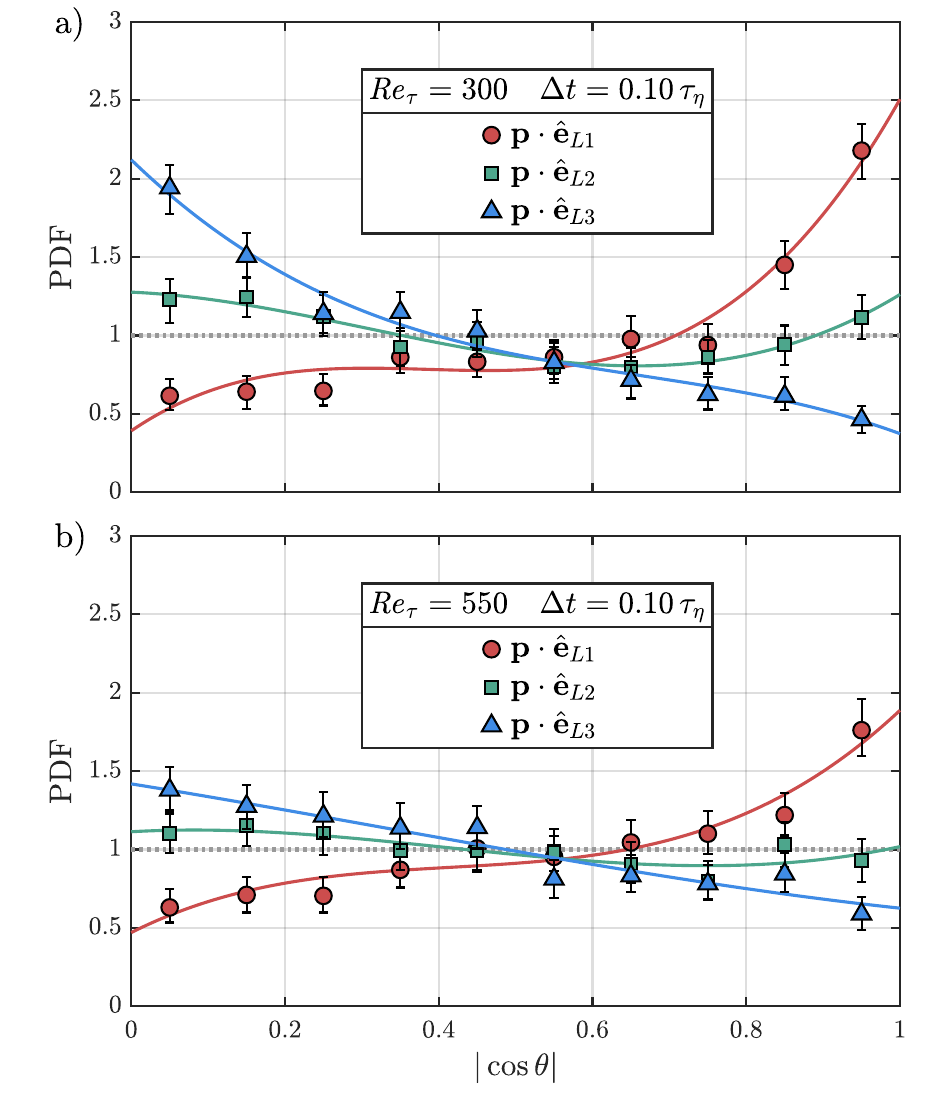}
\caption{ Measured probability density functions of $|\cos\theta|$, where $\theta$ denotes, depending on the eigenvector considered, the angle between the fibre orientation vector $\mathbf{p}$ and the eigenvectors $\hat{\mathbf{e}}_{L1}$, $\hat{\mathbf{e}}_{L2}$, and $\hat{\mathbf{e}}_{L3}$ of the left Cauchy--Green deformation tensor $\mathbf{C}^{(L)}$, for (a) $\mathit{Re}_{\tau}=300$ and (b) $\mathit{Re}_{\tau}=550$. The eigenvectors $\hat{\mathbf{e}}_{L1}$, $\hat{\mathbf{e}}_{L2}$, and $\hat{\mathbf{e}}_{L3}$ correspond to the directions of maximum, intermediate, and minimum finite-time stretching over the integration interval $\Delta t = 0.10\,\tau_{\eta}$, respectively. Symbols denote experimental results obtained in the central region of the channel, while solid lines represent third-order polynomial fits, included to highlight the trends of the measured distributions. Error-bars represent confidence intervals at 99\% confidence level.}
    \label{figure4}
\end{figure}
Thus, Figure~\ref{figure4} reports the measured PDF of $|\cos\theta|$, where $\theta$ denotes, successively, the angle between the fiber orientation vector $\mathbf{p}$ and the eigenvectors $\hat{\mathbf{e}}_{L1}$, $\hat{\mathbf{e}}_{L2}$, and $\hat{\mathbf{e}}_{L3}$ of the left Cauchy--Green deformation tensor.  For both Reynolds numbers considered, the strongest preferential alignment is observed with the leading eigenvector $\hat{\mathbf{e}}_{L1}$, corresponding to the direction of maximum stretching over the interval $\Delta t$. This is evidenced by the pronounced increase of the PDF of $|\mathbf{p}\!\cdot\!\hat{\mathbf{e}}_{L1}|$ toward unity. The probability density associated with the weakest stretching direction $\hat{\mathbf{e}}_{L3}$ is largest near $|\cos\theta|=0$ and decreases monotonically, showing that fibers strongly avoid alignment with this direction. The statistics relative to the intermediate direction $\hat{\mathbf{e}}_{L2}$ remain comparatively flat and close to the random reference level, with only weak departures from uniformity. Together, these trends confirm that, over the relatively short time window considered, fibers preferentially align with the dominant principal deformation direction and tend to remain perpendicular to the third eigenvector $\hat{\mathbf{e}}_{L3}$, associated with the smallest principal stretch ratio.
A comparison between the two Reynolds numbers further shows that the alignment with $\hat{\mathbf{e}}_{L1}$ is stronger at $\mathit{Re}_{\tau}=300$ than at $\mathit{Re}_{\tau}=550$, while the distribution  of $\hat{\mathbf{e}}_{L3}$ is also more pronounced at the lower Reynolds number. As already noted for Figure~\ref{figure3}, this attenuation at higher Reynolds number is presumably associated with stronger finite-size filtering effects.

\subsection*{\fontfamily{phv}\fontsize{9}{11}\bfseries\selectfont Conclusions}
We presented a combined experimental and numerical study of the preferential alignment of Kolmogorov-size fibers in turbulent channel flow at $\mathit{Re}_{\tau}=300$ and $550$, focusing on the channel core. The experimental velocity-gradient measurements were first validated by recovering the canonical alignment of vorticity with the intermediate strain-rate eigenvector and its perpendicularity to the most compressive direction, in close agreement with DNS. The fibers were then found to align most strongly with the local vorticity direction, while exhibiting weaker but still non-random alignments with the strain eigendirections. Analysis based on the left Cauchy--Green tensor further showed that fiber orientation is more closely linked to finite-time Lagrangian deformation than to the instantaneous Eulerian strain eigenframe, with the strongest alignment occurring along the leading principal stretching direction. Overall, the agreement with DNS supports the robustness of the present methodology. The weaker alignment signatures observed at the higher Reynolds number suggest increasing finite-size filtering effects. Future investigations will address the coupling between finite-size fibers and multiscale turbulent deformation, including also the near-wall region.

\vspace{22pt}

% \section*{\fontfamily{phv}\fontsize{9}{11}\bfseries\selectfont REFERENCES}
\bibliographystyle{tsfp}
\bibliography{tsfp}

\end{document}